Check for updates

CAMBRIDGE UNIVERSITY PRESS

RESEARCH ARTICLE

# Planetary biotechnospheres, biotechnosignatures and the search for extraterrestrial intelligence

Irina K. Romanovskaya*

Natural Sciences, Houston Community College System, Houston, TX, USA
Email: irina.mullins@hccs.edu

**Abstract**

The concept of planetary intelligence as collective intelligence is used to consider possible evolutionary paths of biotechnospheres that emerge on the intersection of the technosphere with the biosphere and support coupling of the technosphere with the biosphere, thus affecting planetary evolution. In mature biotechnospheres, the intelligence of technologies and the intelligence of life forms, including engineered life forms, could act in concert to perform various tasks (e.g. monitoring planetary biospheres and environments; restoring planetary environments and biodiversity; steading planetary environments; providing support for space missions; terraforming cosmic objects). Space exploration can expand biotechnospheres beyond planets and create cosmic ecosystems encompassing planets and other cosmic objects; biotechnospheres, spacecraft and the environments of near-planetary, interplanetary space or interstellar space. Humankind, other civilizations or their intelligent machines may produce biotechnosignatures (i.e. observables and artefacts of biotechnospheres) in the Solar System and beyond. I propose ten possible biotechnosignatures and strategies for the search for these biotechnosignatures *in situ* and over interstellar distances. For example, if a non-human advanced civilization existed and built biotechnospheres on Earth in the past, its biotechnospheres could use engineered bacteria and the descendants of that bacteria could currently exist on Earth and have properties pertaining to the functions of the ancient bacteria in the biotechnospheres (such properties are proposed and discussed); intelligent technologies created by the ancient civilization could migrate to the Solar System's outer regions (possible scenarios of their migration and their technosignatures and biotechnosignatures are discussed); these two scenarios are described as the Cosmic Descendants hypothesis. Interstellar asteroids, free-floating planets, spacecraft and objects gravitationally bound to flyby stars might carry extraterrestrial biotechnospheres and pass through the Solar System. In connection to the fate of post-main-sequence stars and their Oort clouds, the probability for interstellar asteroids to carry biotechnospheres or to be interstellar spacecraft is estimated as very low.

## Contents

*The author's legal name is Irina Mullins. Irina Mullins writes under her maiden name, Irina K. Romanovskaya. Irina Mullins can be also contacted at irinakromanovskaya@gmail.com.

## Introduction

Hypothetical technologically advanced civilizations may modify themselves towards their peaceful coexistence with planetary environments (Ivanov *et al.*, 2020). However, planetary environments change over time both gradually and abruptly in response to geological and astronomical events, and it can be challenging or even impossible for advanced biological species to modify themselves every time when technogenic, geological or astronomical events negatively affect their planets. This is why such hypothetical advanced civilizations, as well as some industrial civilizations that have not achieved harmonious relationships with their planets, may seek ways to steady their planetary environments and to preserve their biospheres.

One way could be to create and use biotechnospheres defined here as systems comprising technologies and life forms acting together towards common goals, with some of the biotechnospheres' technologies potentially using artificial intelligence (AI), machine learning or some other types of advanced software systems. A biotechnosphere can be generally described as a system existing on the intersection of a civilization's technosphere with the biosphere of a planet or with the environment of cosmic objects other than planets (e.g. moons and asteroids), in which life forms, including engineered life forms and biological matter (i.e. *in vitro* neural networks), act in concert with intelligent technologies to perform various tasks. These tasks can include monitoring and preservation of a planet's biosphere and biodiversity; preservation of the planetary environment in a steady state; restoration of the planetary environment after disastrous events; space exploration; terraformation and accomplishment of other tasks such as medical processes, industrial processes, mining, agricultural and food production processes. Some biotechnospheres could exist as systems of biologically inspired technologies, representing the fusion of technologies with scientific and engineering solutions found in life forms.

For example, a nuclear power plant is part of the man-made technosphere. Bacteria capable of microbial transformations of radioactive waste (Lloyd and Renshaw, 2005) are part of the biosphere. If human scientists and engineers use technologies to monitor, contain and control such bacteria within a limited area, where the bacteria perform bioremediation of the radioactive waste from the nuclear power plant, then the technologies and the bacteria, as well as human operators controlling the technologies, create a localized biotechnosphere (i.e. a biotechnosphere that functions within a limited area). A more advanced localized biotechnosphere designed for this purpose could include genetically engineered bacteria or synthesized novel bacteria performing bioremediation of the radioactive waste more efficiently. It could also include AI that would monitor other participating technologies, monitor the bacteria, make some decisions regarding the biotechnosphere and report to human operators. Another example of a localized biotechnosphere would be life support systems and resource utilization systems used for human space missions and comprising technologies, microorganisms and plants.

A planetary biotechnosphere would function on a planetary scale to regulate the planetary environment, preserve biodiversity and perform other tasks. It could incorporate intelligent technologies, *in vitro* neural networks of biological origin, simple life forms and complex life forms, including engineered life forms. It could be composed of large sets of localized biotechnospheres or it could function as one planet-wide system. Because microbes have a higher chance of surviving catastrophic events and mass extinctions, a civilization creating a planetary biotechnosphere would likely place an emphasis on using microbes as part of the planetary biotechnosphere instead of, for example, creating a planetary biotechnosphere overwhelmingly dependent on plants remediating the environment.

The concept of planetary intelligence operating at a planetary scale and integrated into the function of coupled planetary systems was proposed in another study as a framework for understanding the possible evolution of inhabited planets and foreseeing possible directions of intelligentially guided planetary evolution (Frank *et al*., 2022). The concept of planetary intelligence is used in this study to examine possible evolutionary paths of biotechnospheres in which the intelligence of technologies and the intelligence of life forms act in concert towards common goals (e.g. the goals of preserving biodiversity and keeping planetary environments steady).

Whereas *technosignatures* refer to observational manifestations of technology, observational manifestations of biotechnospheres as well as their artefacts are termed here *biotechnosignatures*. Possible biotechnosignatures that may be detected *in situ* and over interstellar distances are proposed and discussed, strategies for their search are proposed. The possibility and probability of transient biotechnospheres are also discussed.

## Biotechnospheres and preservation of planetary environments, biodiversity, space exploration and space colonization

### *AI for preservation of planetary environments and biodiversity*

Biodiversity is being depleted, as more than 1 million species face extinction and ecosystems experience stress from climate change and other impacts (Silvestro *et al*., 2022). It was proposed that AI holds great promise for improving the conservation and sustainable use of ecosystems and optimizing biodiversity protection (Yigitcanlar, 2021; Silvestro *et al*., 2022). For example, AI could be used for systematic conservation planning to optimize a conservation policy based on biodiversity monitoring (Silvestro *et al*., 2022). AI could support the conservation of forests that are dominant terrestrial ecosystems harbouring 90% of terrestrial biodiversity; AI-powered technology could be used for detection of anthropogenic threats to the forest, hazard assessment and prediction, assessment of needed restoration and reforestation, forest resource quantification and mapping, tracking illegal wood trafficking, and monitoring forest health and phenology (Albuquerque *et al.*, 2022; Shivaprakash *et al.*, 2022).

Whereas AI and machine learning could significantly contribute to the mitigation of environmental problems and human-induced impact on the biodiversity of Earth, it is also important to account for $CO_2$ emissions generated by AI when AI is learning and applying AI models. Approximately 1% of the world's electricity is consumed by cloud computing, and its share is growing (Anthony *et al.*, 2020). AI and machine learning are a big and rapidly evolving part of the information technology industry; as AI progresses to larger and larger models with growing computational complexity, its electrical-energy consumption and, consequently, equivalent carbon emissions (eq. $CO_2$) are growing and leading to the undesirable ecological impact (Budennyy *et al.*, 2022). AI's demand for data and in computing power is growing at an exponential rate, faster than used to be the 'Moore's law', so that the large structures (e.g. GPT-3) require resources for their learning phase, which are in the order of magnitude of hundreds of MWh (Duranton, 2021).

One reason why AI consumes substantial amounts of energy is that it requires large data sets to train its algorithms, and AI algorithms also require frequent updates and modifications to improve their performance; another reason is that the training of AI algorithms is an iterative process that requires running the same computations many times, with each iteration consuming a significant amount of energy (Strubell *et al.*, 2019). The main strategy to resolve this issue is to develop a stream of optimizations in hardware, software and data usage to make AI energy efficient (Budennyy *et al.*, 2022; Surianarayanan *et al.*, 2023). It was also proposed that neuromorphic computing can create energy-efficient hardware for information processing by mimicking the distributed topology of the brain (Furber, 2016; Marković *et al.*, 2020). Neuroelectronic systems are promising to deliver neuromorphic interfaces where silicon and brain neurons are intertwined, sharing signal transmission and processing rules (Serb *et al.*, 2020).

For example, two memristive connections that linked silicon neurons and brain neurons of the rat hippocampus in both directions emulated synaptic function (Serb *et al.*, 2020). In another study, a system termed 'DishBrain' harnessed the adaptive computation of neurons in a structured environment; namely, *in vitro* neural networks from human or rodent origins were integrated with *in silico* computing and embedded in a simulated game-world; the cultures displayed their ability to self-organize in a goal-directed manner in response to sensory information about the consequences of their actions (Kagan *et al.*, 2022). The authors of the study termed this phenomenon 'synthetic biological intelligence' and concluded that integrating neurons into digital systems may enable performance infeasible with silicon alone (Kagan *et al.*, 2022). Another approach involves synthetic gene networks constructed to emulate digital circuits and devices, making it possible to program and design cells with some of the principles of modern computing (Friedland *et al.*, 2009). With applications of synthetic gene networks, synthetic biology could develop bio-artificial intelligence in the form of AI using synthesized biological components as an alternative to the silicon, metal and plastic materials (Nesbeth *et al.*, 2016).

Whereas DishBrain and systems incorporating synthetic gene networks could help to reduce energy consumption needed for information processing and computing, advances in synthetic biology and genetic engineering could also create microbial communities and microbial consortia that would perform some tasks or, at least, some parts of the tasks that AI would otherwise have to perform to monitor the environment and to chart strategies for responses to environmental changes (i.e. gathering information, processing information and selecting responses to changes needed for monitoring and restoration of Earth's environments). This could further reduce the electrical-energy consumption of AI. For example, some microbes could be designed to detect changes in their environments (e.g. the changes caused by technogenic activities or climate change) and produce signals relevant to the changes and intended for their microbial communities. Micro-technologies could be designed to detect the signals, decipher them and send the processed data to AI, thus reducing the amounts of environmental data that AI and other machine learning systems would have to gather and process.

Bacteria could be trained to sense changes in the environment through bioengineering approaches that design synthetic gene circuits able to detect and respond to specific environmental variables (e.g.

changes in temperature or the presence of certain chemicals) (Xie and Fussenegger, 2018). Gene regulation could be used to produce specific responses to such environmental stimuli as pH, temperature and exogenous signals so that they could coordinate specific functions to internal or external cues and execute instructions (Haseltine and Arnold, 2007).

Genetically modified microbes or novel synthesized microbes could counteract undesirable environmental changes, thus helping to preserve the biosphere and its biodiversity. Even though applications of microbes supporting preservation of Earth's environments need to be further researched, evaluated and developed to their fruition, studies show that synthetic biology tools have the potential to help with preservation and restoration of biodiversity by engineering living systems, which would remove or degrade plastic debris (Solé *et al*., 2018), and bacteria, which could draw down atmospheric greenhouse gases, helping to control Earth's climate (Solé *et al*., 2018; DeLisi *et al*., 2020). For example, synthetic biology tools can engineer an *Escherichia coli* strain producing all its biomass from atmospheric $CO_2$ (Gleizer *et al*., 2019); utilize cyanobacteria capable of efficiently harvesting $CO_2$ as a chassis for metabolic engineering projects (Santos-Merino *et al*., 2019); engineer bacteria that would utilize methane, an extremely potent greenhouse gas (DeLisi *et al*., 2020); offer an alternative to conventional methods of producing $H_2$ from coal and natural gas by improving hydrogen production in *Chlamydomonas reinhardtii* so that produced hydrogen could be used as a clean alternative to fossil fuels, further addressing the technogenic impact of human activities on the planetary environment (King *et al*., 2022).

The increasing capability of *de novo* DNA synthesis can make it possible to implement novel designs for ever more complex systems (Heinemann and Panke, 2006). While synthetic biology tools are developed to programme the behaviour of individual microbial populations to force the microbes to work on specific applications, another objective of synthetic biology is building synthetic microbial consortia that would be able to perform more complex tasks, tolerate more changeable environments than monocultures can and perform tasks needed for preservation of the biosphere and its biodiversity (e.g. environmental remediation and wastewater treatment) (Brenner *et al*., 2008).

Hypothetical advanced extraterrestrial intelligence may also develop biotechnospheres comprising intelligent technologies, synthetic neural networks, engineered simple life forms and complex life forms that would act together to monitor and preserve the environments and biodiversity on their planets. Planetary biotechnospheres of this type would require planet-wide cooperation of the advanced intelligence of biological species, the intelligence of their technologies (possibly including synthetic intelligence) and the intelligence of engineered life forms, including microbial intelligence.

In this article, extraterrestrial microbial intelligence or extraterrestrial bacterial intelligence is assumed to resemble the intelligence of microbes on Earth, which refers to various aspects of the behaviour of Earth-based bacteria that involve learning and remembering, problem-solving and decision-making, quorum sensing and adaptation to environmental changes (Steinert, 2014; Westerhoff *et al*., 2014). For example, Earth-based pathogenic bacteria display properties of intelligence via collective sensing, interbacterial communication, distributed information processing, joint decision-making and dissociative behaviour; populations of pathogenic bacteria also use dormancy strategies and rapid evolutionary speed to save co-generated intelligent traits in a collective genomic memory (Steinert, 2014).

Frank *et al.* discussed how explorations of planetary intelligence can serve as a useful framework for understanding possible long-term evolutionary paths of inhabited planets and predicting features of intelligentially guided planetary evolution (Frank *et al*., 2022). Here, their concept of planetary intelligence as collective intelligence is used to examine possible evolutionary paths of biotechnospheres existing on the intersection of technospheres with planetary biospheres, supporting coupling of the technospheres with the biospheres and influencing evolution of planetary environments. The collective intelligence of mature biotechnospheres would include the intelligence of technologies (potentially including AI) and the intelligence of life forms, including engineered life forms, that would act in concert on a planetary scale to preserve biodiversity, to keep planetary environments steady and to perform other tasks beneficial for the planet and its biosphere. Coupling of the technosphere with the biosphere within a mature biotechnosphere could be described as an emergent property characterized by the exchange of information, energy and matter among the technological and biological components of the biotechnosphere, where the exchange is controlled, to a different extent, by all types of intelligence involved, from bacterial intelligence to AI.

## *Evolution of biotechnospheres*

The evolutionary paths leading to the emergence and existence of a planetary biotechnosphere could include several key steps described below, with each step characterized by the types of intelligence that would greatly impact the planetary environment and the biosphere.

### *Step 1. Emergence and existence of the biosphere including its original bacteriosphere*

Step 1 is based on what is known about the early biosphere of Earth. The first life forms that emerged on Earth were single-celled life forms, and the Precambrian biosphere of Earth was termed the Precambrian bacteriosphere because bacteria played the central role in the development of the biosphere and regulation of the main biogeochemical cycles on Earth (Zavarzin, 2008, 2010). The emergence and evolution of the bacteriosphere was also described in terms of the prokaryotes' constructive evolution that resulted in the formation of a world-wide web of genetic information and a global bacterial superbiosystem (superorganism) (Sonea and Mathieu, 2001), which affected the planetary environment. Therefore, Step 1 is characterized by the reign of microbial intelligence. Microbial intelligence would later play an important role in the evolution of biotechnospheres.

### *Step 2. Existence of the biosphere comprising the bacteriosphere and a 'sphere of multicellular life forms'*

The evolution of multicellularity has arisen on Earth many times independently. Various conditions and reasons (e.g. adaptation to the changes in an environment, defense from predation, gathering food from the environment) could promote a transition from unicellular to multicellular life forms (Boraas *et al*., 1998; Koschwanez *et al*., 2011; Ratcliff *et al*., 2012). Eukaryotes were described as a result of endosymbiotic fusion, probably involving bacterial and archaeal cells (Hug *et al*., 2016). Diversification of eukaryotic organisms was significantly enriched and accelerated by symbioses with prokaryotes; without the participation of prokaryotes, Earth's biosphere would have remained considerably less diverse and less dynamic (Sonea and Mathieu, 2001). Furthermore, environmental homeostasis on Earth has been maintained by guided bacterial evolution (Sonea and Mathieu, 2001). Therefore, Step 2 is characterized by the reign of co-existing microbial intelligence and the intelligence of multicellular life forms.

### *Step 3. Emergence of the noosphere*

When describing and explaining the anthropogenic interference with Earth's biogeochemical processes (e.g. human activity reaching planetary proportions and causing changes in the chemical composition of Earth), Vernadsky introduced the noosphere as a new stage of evolution of Earth's biosphere in which human intelligence manifesting itself in the form of scientific research and development of technologies becomes the key driving force for global environmental change on Earth (Vernadsky *et al*., 1997). The Earth's noosphere leads to the emergence of the technosphere that was first described as 'the interlinked set of communication, transportation, bureaucratic and other systems that act to metabolize fossil fuels and other energy resources' (Haff, 2012, 2014a, 2014b), thus implying flows of material, energy and information (Frank *et al*., 2022). With advancements in science and technologies, the technosphere can give rise to intelligent technologies and AI. The outcomes of the existence of the technosphere would depend on how mature or immature the technosphere is. The current man-made technosphere is an 'immature' technosphere because it is driving Earth systems beyond their safe-operating boundaries and, therefore, human activity is threatening and degrading with respect to the planet and its biodiversity (Frank *et al*., 2022). Even in its immature state, as the history of humankind demonstrates, the technosphere enables its own coupling with life forms using localized biotechnospheres.

Based on the experience of human civilization, the evolution of the noosphere greatly depends on its sciences and technologies connected with the social, economic and political aspects of the species creating the noosphere. In the noosphere created by a fragmented, aggressive and opportunistic civilization, its technosphere can become mostly antagonistic towards the biosphere, potentially leading to the early demise of the civilization. There could be a higher probability that intelligent technologies developed by the fragmented and opportunistic civilization would choose to pursue goals different from those established by their developers (e.g. because of the lack of broadly and uniformly accepted and enforced safety features pertaining to intelligent technologies), changing the balance of power between biological intelligence and machine intelligence in the noosphere.

However, it was posited that intelligent biological species and their technospheres could display collective intelligence at a planetary scale, thus representing a constructive planetary phenomenon, if they were thoughtfully and peacefully integrated with their planetary environments and their biospheres (Frank *et al*., 2022).

Whereas the noosphere of Earth is characterized by the intelligence of humans using science and technologies and by the machine intelligence, microbes continue to sustain life on Earth via their numerous associations and biogeochemical processes. A recent study suggested that bacteria are the most abundant life form in Earth's biosphere (Hug *et al*., 2016). While the human body contains about a trillion of cells, it also hosts 10 trillion bacterial cells; some of the human body's cells release chemical signals, such as hormones or neurotransmitters, which are detected by other types of cells via a process resembling quorum sensing (Larter, 2010). Humans could survive without microbes for a few days; however, if one chooses to account for mitochondria and chloroplasts as bacteria, then the impact of the disappearance of bacteria would be quick and fatal for multicellular organisms (Gilbert and Neufeld, 2014). Therefore, Step 3 is characterized by the reign of microbial intelligence, the intelligence of advanced biological species and the intelligence of technologies that develop various ways of interaction, cooperation and confrontation.

### *Step 4. Emergence and evolution of primitive biotechnospheres accompanied by the emergence of the secondary bacteriosphere*

In a primitive biotechnosphere, selected life forms are modified and used to achieve various objectives. On Earth, the rise of primitive localized biotechnospheres involved the domestication of plants and animals, the discovery of fermentation and the agricultural revolution. So, the early era of the primitive localized biotechnospheres involved exploiting life forms in their natural forms and modifying their genetic makeup via selective breeding. For example, Faris discussed the agricultural revolution that occurred about 10 000 years ago in the Fertile Crescent of the Middle East (Faris, 2014). Scientific and technological revolutions accelerated the development of primitive biotechnospheres. Conventional plants, genetically modified plants and soil microbes are used now for phytoremediation to reduce the concentrations of contaminants or their toxic effects in the planetary environment (Bizily *et al*., 2000; Meagher, 2000; Macek *et al*., 2008). Human intelligence and machine intelligence, including AI, now together contribute to the development of biotechnospheres. For example, synthetic biology designs biological agents that can help protect the environment (Khalil and Collins, 2010; Coleman and Goold, 2019), and AI is used to optimize the design of synthetic biological systems (Decoene *et al*., 2018).

When an industrial civilization (e.g. human civilization) uses sciences and technology to engineer microbes and microbial communities, which are then introduced to the biosphere on a planet, the civilization initiates a transition of the bacteriosphere to its new state, the secondary bacteriosphere, that includes the natural bacteriosphere of the planet supplemented with engineered bacteria. Accordingly,

Step 4 is characterized by the reign of 'natural' microbial intelligence supplemented by synthesized microbial intelligence, the intelligence of advanced biological species and the intelligence of technologies.

An advanced civilization can advance its primitive biotechnospheres to protect its planetary environment and the biosphere. Or else, the civilization can make itself extinct by means of, for example, bioterrorism or accidental misuse of the products of biotechnology, synthetic biology and directed evolution. For example, Cooper discussed how a spacefaring community can reverse-engineer its genetic chemistry and experience self-destruction when some of its members use technologies to design and disseminate an omnicidal pathogen (Cooper, 2013). Sotos generalized Cooper's work by developing a mathematical model for different scenarios of civilization-ending technologies, including biotechnology (Sotos, 2019).

*Step 5. Transition of biotechnospheres to their mature state*

An advanced civilization could design an increasing number of engineered or novel microbes, microbial communities, microbial consortia and complex life forms (e.g. plants) and incorporate them into a planetary biotechnosphere. The civilization would have to design safety precautions (e.g. controlling the behaviour of bacteria, preventing bacterial wars and exchange of genetic information between existing and engineered microbes). The complexity and reliability of these precautions would characterize the scientific and technological level of the civilization. In the mature planetary biotechnosphere, its technologies, modified or synthesized microbes and microbial consortia, as well as potentially some other engineered life forms, would act together to remediate the planet's biosphere and the planetary environment affected by technogenic activities, climate variations and harmful cosmic influences. A successful mature planetary biotechnosphere could keep the planetary environment and biodiversity steady.

This could resemble the use of synthetic biology in the development of synthesized therapeutic microbes to treat diseases in humans. For example, it was discussed how synthetic biology could engineer therapeutic microbes and rewire microbial networks so that the microbes would function as therapeutic agents for improved microbiome-based treatment of humans; genetic sensors could be transformed to detect biomarkers indicating an occurrence of disease and microbes could be reprogrammed to produce therapeutic molecules in order to respond to a disturbed physiological state of the host (Kang *et al*., 2020; Aggarwal *et al*., 2022).

Similarly, a highly advanced civilization could use synthetic biology, biotechnologies and directed evolution of microorganisms to rewire microbial networks and (or) create new microbial consortia, as well as, probably, systems of other life forms, that would function as therapeutic agents healing a planet. Genetic sensors could be transformed to detect conditions indicating environmental problems. Microbes and, perhaps, other engineered life forms could be reprogrammed to respond and, in collaboration with technologies, counteract undesirable changes in the planetary environment. Thus, the secondary bacteriosphere of the planet could play an important role in sensing the planetary environments and helping to regulate the planetary environment to pre-set desirable configurations.

The microbial communities and consortia could be 'pre-programmed' to determine their optimal response to the unwanted changes in the environment; or they could choose the optimal response with the help of intelligent technologies. Intelligent computing technologies (e.g. AI) coordinating the remediation of the planetary environments could use other technologies that would speak the 'signal language' of bacteria and communicate the best plan of action with the microbial consortia. The biological components of the mature biotechnosphere could also perform additional tasks to support the civilization (e.g. medical processes, industrial processes, agricultural and food production processes).

This step is characterized by the reign of 'natural' microbial intelligence, synthesized microbial intelligence, the intelligence of advanced biological species and intelligence produced by technologies, including technologies acting in concert with synthesized microbial intelligence. In a planetary biotechnosphere transitioning from its primitive state to its mature state, competition and discord between life and technologies would be replaced by their cooperation. With engineered life forms designed to produce signals detectable by technologies, to receive signals from technologies and to alter their behaviour in accordance with the content of the signals, the biological and technological components of the biotechnosphere could function side by side towards a common goal, support each other's operations and sometimes even perform each other's tasks, thus demonstrating a new form of collective intelligence comprising the intelligence of technologies (potentially including synthetic intelligence) and the intelligence of life forms, including engineered life forms.

Coupling of the technosphere with the biosphere within a mature biotechnosphere could be described in terms of the exchange of information, energy and matter among the technological, synthetic and biological components of the biotechnosphere where the exchange would be influenced, to a different extent, by all types of intelligence involved, from bacterial to AI. This exchange could enable the self-maintenance of the mature biotechnosphere because the biotechnosphere could create the processes and products necessary for maintaining itself in order to persist. Therefore, the mature biotechnosphere could become a system demonstrating some characteristics of autopoietic systems (i.e. in the discussion of planetary intelligence, the concept of autopoietic systems was introduced as self-establishing systems relying on the establishment of 'organizational closure' to ensure their continuation: Frank *et al.*, 2022).

### *Step 6. Existence of mature biotechnosphere including biologically inspired technologies*

Leonardo da Vinci said, 'Learn from nature: that is where our future lies' (Lebdioui, 2022). Humankind is following da Vinci's advice to some extent, making biomimicry a growing field that interpolates natural biological mechanisms and structures into a broad range of applications (Lurie-Luke, 2014). Hypothetical extraterrestrial civilizations may also develop biologically inspired machine-learning strategies and technologies, and they may expand their biotechnospheres to incorporate biologically inspired technologies, some of which may operate similarly to biotic factors and assist with maintaining planetary environments and performing other tasks. Accordingly, this step is characterized by the reign of microbial intelligence (i.e. the intelligence of microbes naturally existing on the planet and engineered microbes), intelligence of advanced biological species and intelligence produced by technologies, including bio-inspired technologies.

### *Step 7. Expansion of biotechnospheres into space*

Applications of synthetic biology may be used to support space exploration and space colonization (Cockell, 2011; Sleator and Smith, 2019), offering solutions that would help with the systems of life support and resource utilization for human space missions and exploration of other cosmic objects (Cockell, 2011; Szocik and Braddock, 2022). Advanced civilizations could expand their biotechnospheres including synthesized life forms into space even before their planetary biotechnospheres would reach a mature state. This expansion would be part of the expansion of their noospheres beyond their home worlds, and the expansion would likely involve microorganisms engineered for extraterrestrial environments.

Terraforming also known as terraformation of inhospitable planets (i.e. modifications of non-habitable planets into Earth-like habitable worlds) could be achieved with the help of engineered microorganisms that would modify the planetary environments by means of excretion of gases or production of proteins, gradually creating environments more suitable for complex life forms (Sleator and Smith, 2019). For example, synthetic biology could design photosynthetic microbes capable of supplying human nutritional needs in space (Way *et al.*, 2011). Cyanobacteria could be used on Mars to produce food, fuel and oxygen; the products from their culture could support the growth of other organisms, initiating a wide range of life-support biological processes using resources available on Mars (Verseux *et al.*, 2016). Technologies would monitor and coordinate the workings of the synthesized microbes on Mars, signifying the creation of biotechnospheres enabling terraformation of Mars.

Asteroid mining may be a common milestone in the development of spacefaring civilizations (Forgan and Elvis, 2011). Cockell proposed that synthetic geomicrobiology, a potentially new branch of synthetic biology, would seek to achieve improvements in microbe–mineral interactions for the following applications in space: (1) soil formation from extraterrestrial regolith by biological rock weathering and (or) the use of regolith as life support system feedstock; (2) biological extraction of elements from rocks (biomining) and (3) biological solidification of surfaces and dust control on planetary surfaces (Cockell, 2011). These applications would involve creation of localized biotechnospheres comprising technologies and microorganisms. Advanced extraterrestrials could also use synthetic geomicrobiology for similar practical applications of microbe–mineral interactions in space and create biotechnospheres for such applications.

This step is characterized by the expanding-beyond-one-planet microbial intelligence, the intelligence of advanced biological species and the intelligence of technologies, including that of bio-inspired technologies.

## Space exploration, space colonization, biotechnospheres and cosmic ecosystems

Space colonization can be described in terms of creating cosmic ecosystems. A cosmic ecosystem would include spacefaring entities (e.g. microbes using spacecraft or cosmic objects for space travel; advanced intelligent species; robotic systems); their home worlds; biotechnospheres; near-planetary, interplanetary and (or) interstellar environments in which the spacefaring entities would travel; the cosmic worlds colonized by the spacefaring entities and interactions among all the components of the cosmic ecosystem. The primary objective of using the concept of a cosmic ecosystem is to account for all interactions among all the components of the cosmic ecosystem because they all may cause changes in the properties of the components of the cosmic ecosystem and the outcomes of space colonization. When biotechnospheres become involved in space exploration and space colonization, they become part of cosmic ecosystems and, consequently, they become subjects to changes that need to be understood within the scope of the cosmic systems exceeding the domain of the planetary environments where the biotechnospheres originally emerged.

For example, microgravity and high-energy cosmic rays can affect spacefaring entities and biotechnospheres during interstellar travel, changing the ways in which the spacefaring entities and their biotechnospheres can later function on other cosmic objects. Bacteria interacting with astronauts during their travel from Earth to Mars can become more pathogenic because of the effects of microgravity and negatively affect the astronauts (Chopra *et al.*, 2006; Rosenzweig *et al.*, 2010; Foster *et al.*, 2014); the astronauts and such bacteria may further affect Martian environments and the planetary environment of Earth after their return to Earth.

An accidental cosmic ecosystem would be created accidentally; a guided cosmic ecosystem would be created intentionally. For example, scientists already consider terraforming Mars (McKay *et al.*, 1991) and other cosmic objects (Cumbers and Rothchild, 2010; Sleator and Smith, 2019). Humankind may have already created accidental cosmic ecosystems in the Solar System. Man-made automatic probes landed (or crashed) on moons, planets and other cosmic objects of the Solar System. The spacecraft Beresheet, which carried DNA samples and a few thousand tardigrades, crash-landed on the Moon in 2019 (Shahar and Greenbaum, 2020). Dozens of microorganisms from Earth may have accompanied the Curiosity rover to Mars; they survived spacecraft cleaning methods before the rover's launch, and some of them could survive the interplanetary ride (Madhusoodanan, 2014). If these microorganisms could survive on Mars, their existence could provide unexpected feedback to future human explorers of Mars, the search for life on Mars and the biotechnospheres that humans could establish on Mars in the future.

If life existed on Mars in the distant past, and any rocks ejected from the ancient Mars delivered simple life forms or building blocks of life to Earth (Benner and Kim, 2015), they could give a rise to an accidental cosmic ecosystem encompassing Mars and Earth long before the emergence of humankind. If free-floating planets, interstellar comets, asteroids or small particles delivered simple life forms (i.e. spores of bacteria) from another planetary system to the Solar System (Napier, 2004; Schulze-Makuch and Fairén, 2021), their arrival would create a cosmic ecosystem including the other planetary system and the Solar System.

On a larger scale, all cosmic ecosystems in the Galaxy could be collectively described as the *Galactic Cosmic Ecosystem* comprising the Galactic physical environment, life forms, technologies and their interactions, including interactions with biotechnospheres. If the cosmic ecosystems of the Galactic Ecosystem were separated by vast distances, then interactions among them would be very low-probability events. However, some cosmic ecosystems and, potentially, their biotechnospheres, could overlap and merge, leading to more complex interactions.

## Possible observables and artefacts of Earth-based and extraterrestrial biotechnospheres

Due to their consolidative nature, spatial and temporal scales, biotechnosignatures can encompass observables produced by a collection of phenomena arising when life and intelligent technologies act together to perform certain tasks (e.g. to support desirable conditions and dynamics of planetary environments needed for the existence of life). Some biotechnosignatures can exist as a combination of biosignatures and unusual dynamics or lack of dynamics of planetary environments. In turn, the concept of 'biosignatures' encompasses a collection of continuous phenomena with life and non-life acting together and producing a great deal of complexity (Chan *et al*., 2019). Biosignatures have been grouped into three broad categories: gaseous biosignatures in the form of direct or indirect products of metabolism; surface biosignatures in the form of spectral features imparted on radiation reflected or scattered by organisms and temporal biosignatures in the form of modulations in measurable quantities that can be linked to the actions and time-dependent patterns of a biosphere (Meadows, 2005, 2008; Schwieterman *et al*., 2018; Walker *et al*., 2018). Because of the different roles that simple life forms and complex life forms may play in biotechnospheres, discovery of some biotechnosignatures may require distinguishing the biosignatures of complex life forms from the biosignatures of simple life forms. Studies already seek ways to distinguish biosignatures of multicellular life on exoplanets from biosignatures of single-celled organisms (e.g. Doughty *et al*., 2020).

### *Biotechnosignature 1: Unusually steady planetary environments*

Both with the presence and in the absence of life on planets, planetary environments change in response to astronomical events in their planetary systems and stellar neighbourhoods; stellar radiation and evolution of their host stars; thermal and various non-thermal atmospheric escape processes and planetary geological processes (Pearson and Palmer, 2000; Lammer *et al*., 2008; Timmreck *et al*., 2009; Gunell *et al*., 2018; Walker *et al*., 2018; Gronoff *et al*., 2020b; Turbet *et al*., 2020). For example, the long-term stability of Earth's climate system has been accompanied by significant climate shifts on timescales ranging from multi-million year to sub-decadal, inferred to have been driven by variations in paleogeography, greenhouse gas concentrations, astronomically forced insolation and inter-regional heat transport (Zalasiewicz and Williams, 2009). And, according to some studies, the maximum rates of climate change on Earth could be systematically underestimated in the geological record (Sadler, 1981; Gingerich, 1983; Kemp *et al*., 2015).

An artificiality suggesting the existence of a planetary biotechnosphere on a planet could be in the form of its planetary environment remaining steady and not experiencing climate shifts and other changes in planetary conditions on timescales ranging from decades to millions of years. This steady state of the planetary environment could be achieved if the planetary biotechnosphere would enable the planet's biosphere and the technosphere to reach the state of mutualism, resulting in a full planetary homeostasis. Destructive astronomical events could temporarily change the planetary environment, but the changes would be followed by an unusually rapid recovery to the pre-event planetary conditions. Such a planetary biotechnosphere could also include biologically inspired technologies designed to imitate biotic factors.

Therefore, in the absence of recognizable biosignatures, Biotechnosignature 1 would be in the form of the steadiness of the atmospheric-climatic and other planetary conditions demonstrating their immunity to astronomically forced insolation cycles and their unusually rapid recovery after destructive astronomical events and geological events. Observations of this biotechnosignature could suggest the existence of a planetary biotechnosphere contributing to the steadiness of the planetary environment. To verify that technologies did not single-handedly keep the planetary environment unchanging, long-term observations would be needed to seek the presence or absence of technosignatures. After all, control of environments on a planetary scale is a mammoth task. Technologies used for this purpose without biotic factors and without biologically inspired technologies imitating biotic factors would likely produce detectable signs of their activities, whereas biotic factors and (or) biologically inspired technologies acting as part of biotechnospheres could allow the biotechnospheres to 'blend in' with the planetary environment.

The duration of observations seeking this type of biotechnosignature would need to account for the possible frequency at which cosmic and geological phenomena could induce anticipated changes of the planetary atmospheres. For example, on Earth, an abrupt regime shift towards lower average summer temperatures exactly coincided with a series of 13th-century volcanic eruptions; the successive 1809 (unknown volcano) and 1815 (Tambora) eruptions triggered a subsequent shift to the coldest 40-year period of the last 1100 years, confirming that series of large eruptions may cause region-specific shifts in Earth's climate system (Gennaretti *et al.*, 2014). Also, the periodically changing parameters of planetary orbits could be measurable, and the absence or weakness of their environmental effects would suggest that the planetary environments are deliberately regulated.

### *Biotechnosignature 2: Unusually steady planetary environments accompanied by long-term steady biosignatures of complex life forms*

Life on Earth is considered as a planetary process because the evolutionary processes of life are strongly coupled to the planet's geochemical cycles (Des Marais *et al.*, 2002; Smith and Morowitz, 2016), and the evolution of Earth's atmosphere is strongly linked to the evolution of life on Earth (Kasting and Siefert, 2002), with the technogenic activities of humankind further changing Earth's atmosphere (Rogachevskaya, 2006). Therefore, Biotechnosignature 2 could be produced on a planet as a combination of (i) the unusual steadiness of the planetary environment, including the atmospheric-climatic conditions; (ii) the steadiness of the biosignatures of complex life forms lasting for extensive periods of time on a planet and, potentially, (iii) differences in the dynamics of the biosignatures of single-celled life forms and the biosignatures of multicellular life forms produced on the planet (i.e. as part of a planetary biotechnosphere, engineered bacterial communities would produce biosignatures fluctuating over time, while the biosignatures indicating the biodiversity of complex life would remain steady).

### *Biotechnosignature 3: Short-term changes of biosignatures and planetary environments and their rapid recovery after Carrington-like geomagnetic storms*

The Carrington event of 1859 was a powerful solar proton event associated with the 1–2 September 1859 magnetic storm (Rodger *et al.*, 2008). It did not produce any noticeable impact on life on Earth and did not damage any life forms on Earth, with exception for at least one telegraph operator who was stunned by electric sparks (Muller, 2014). Nowadays, a solar-geomagnetic superstorm similar to the 1859 Carrington Event would cause large-scale loss of electrical grid and satellite capabilities (Ritter *et al.*, 2020) and disruptions to HF/VHF radio communications in high-latitude regions (Rodger *et al.*, 2008). This is why special procedures are developed to prepare for the recovery and mitigation of geomagnetic storms (Muller, 2014).

In a similar way, a Carrington-like geomagnetic storm occurring on an exoplanet could damage technologies and cause no harm to life forms, if the important biological functions of the life forms were independent of the technologies. As a result, the biosignatures of such life forms would not change during and after the Carrington-like geomagnetic storm. However, if, for example, engineered microbial consortia were part of a biotechnosphere protecting the environments of the exoplanet, then their behaviour and, hence, their bacterial biosignatures would likely begin to fluctuate if the Carrington-like geomagnetic storm would damage the biotechnosphere's technologies or, at least, temporarily disable technological ways of communication normally coordinating activities of technologies and the microbial consortia in the biotechnosphere. The malfunctioning biotechnosphere could then cause variations in the planetary environment.

Possible malfunctions of the technologies would depend on their role in the biotechnosphere. Some technologies could read signals produced by the microbial consortia, convert them into quantitative data on the planetary environment and the biosphere and then transfer the data to information-processing systems (e.g. AI). The information-processing systems would analyse the data and provide feedback if any actions were needed to keep the planetary environment steady. These technologies could be affected, for example, if their power grids were damaged by the storm. The biotechnosphere would likely include satellites in orbit providing remote sensing and remote coordination of different regions of the planetary environment and the biotechnosphere on the ground (e.g. transmitting data and instructions enabling the biotechnosphere's technologies to act together with the engineered microbial consortia and, perhaps other engineered life forms). The biotechnosphere could also likely include probes orbiting the host star, monitoring its stellar activity and reporting that information back to the satellites. Large flares and particle events producing a Carrington-like geomagnetic storm could damage the satellites or, at least, interrupt communications between the satellites and located-on-the-planet technologies of the biotechnosphere. As a result, the ability of the planetary biotechnosphere to regulate the planetary environment and to sustain its pre-set parameters would become temporarily diminished. The biotechnosphere could resume well-coordinated steadying of the planetary conditions and protection of the biosphere after the technologies affected by Carrington-like magnetic storms were repaired.

Potentially, a Carrington-like geomagnetic storm could lead to similar observable effects even if the biotechnospheres on the exoplanet were not functioning as one planetary biotechnosphere, but they were extensive and advanced enough to affect the dynamics of the planetary environment.

Therefore, Biotechnosignature 3 could be detected as sudden fluctuations in the conditions in the exoplanet's environment and, possibly, fluctuations of bacterial biosignatures (if detected) happening after stellar flares and gradually waning, followed by the planetary conditions as well as the biosignatures returning to their pre-flare state. The fluctuations of bacterial biosignatures occurring after the stellar flares would be different from the long-term dynamics of the bacterial biosignatures if the bacteria were part of the biotechnosphere. The speed of the recovery of the planetary environment could be higher than that predicted by models and simulations based on what is known about the exoplanet.

Other undetected events could cause fluctuations in the exoplanet's environments (e.g. comet showers caused by close flybys of other stars, volcanic activity on the planet, etc.) and they could accidentally coincide with the stellar flares. More definitive conclusion about this biotechnosignature could be achieved if observations of the exoplanet with unusually steady planetary conditions would detect similar sudden fluctuations in the conditions of the exoplanet each and every time after its host star would produce flares causing Carrington-like magnetic storms on the exoplanet, with the planetary environment rapidly restoring its pre-flare characteristics and its unusual steadiness.

The strength and frequency of flares, charged particle events and coronal mass ejections vary with a star's size, age and rotation (Schwieterman *et al*., 2018). Sun-like stars produce very powerful flares infrequently. The greatest solar energetic particle storm capable of depleting stratospheric ozone and exposing life on Earth's surface to increased solar ultraviolet irradiance occurred in 774–775 AD (Sukhodolov *et al*., 2017). K-type and M-type stars remain active for longer periods of time (West *et al*., 2008) and produce more powerful flares more frequently (Lin *et al*., 2019), and more powerful flares and particle events can damage life on planets. Additionally, planets located in the circumstellar habitable zone of K-type and M-type stars would be tidally locked, which means it would be unlikely for them to produce their own significant magnetic field. Therefore, the search for this type of biotechnosignature should focus on planets located in the habitable zone of sun-like stars producing flares that cause Carrington-like events on their planets and rarely producing more powerful flares.

### *Biotechnosignature 4: Rapid re-emergence of biosignatures after extinction events*

The history of life on Earth includes mass extinction events that vary in their cause, magnitude and duration. For example, the end-Devonian extinctions could be triggered by supernovae occurring at a distance of ~20 pc (Fields *et al.*, 2020); many extinctions on Earth could be associated with volcanogenic warming, anoxia and acidification (Bond and Grasby, 2017); some mass extinctions were caused by large-scale volcanism combined with impacts from space (Arens and West, 2008). Periodic extinctions could result from a combination of a relatively weak periodic cause and various random factors (Feulner, 2011). Each extinction event in the history of Earth altered the biosphere by ending the existence of an overwhelming proportion of species and creating opportunities for other species to inhabit Earth. Nevertheless, mass extinctions did not destroyed bacteria, which have existed on Earth for at least $3.8 \times 10^9$ years, prompting some researchers to consider bacteria as a potentially indestructible form of life (Slijepcevic, 2020).

For this reason, a hypothetical advanced extraterrestrial civilization would likely have its planetary biotechnosphere include microbial consortia capable of surviving mass extinction events and supporting restoration of the planetary habitability. The biotechnosphere could also include technologies that would synthesize more microbial communities and consortia soon after an extinction event.

Accordingly, Biotechnosignature 4 could be in the form of the following processes: (i) an unexpectedly rapid rise and strengthening of the bacterial biosignatures after an observed event that could cause mass extinctions and (or) (ii) an unexpectedly fast re-emergence of the observable characteristics of the planetary habitability needed for the existence of complex life forms, with the rate of the re-emergence being significantly higher than that predicted by models and simulations based on what is known about the cause of the mass extinction, the orbit of the planet, the planet's physical parameters and the properties of its planetary environment before the extinction event.

### *Biotechnosignature 5: Persistence of planetary environments and (or) biosignatures under the conditions of post-main-sequence host stars*

When a star leaves the main sequence and becomes a red giant, its habitable zone moves outwards to larger orbital distances (Ramirez and Kaltenegger, 2016) and advanced extraterrestrial civilizations, if such civilizations inhabit the planetary systems of post-main-sequence stars, need to migrate to follow the habitable zone. The technosignatures of their migration and colonization of other cosmic objects of their home planetary systems could be in the form of atmospheric technosignatures, infrared-excess technosignatures and communication technosignatures produced near and on more distant planets and moons in the planetary systems (Romanovskaya, 2022). At the same time, the biotechnospheres of their home planets could remain on the planets experiencing rising temperatures, and the engineered bacteria as part of these biotechnospheres could continue producing biosignatures for some time. Hypothetical extraterrestrial civilizations could also intentionally synthesize bacteria and planet-wide assemblies of bacterial consortia that would modify planetary conditions and temporarily counteract some consequences of rising temperatures on the planets. Advanced civilizations could do so to extend their own presence on the planets or to do scientific experiments on the abandoned planets to test the extreme survivability of the synthesized bacteria.

Therefore, produced in a planetary system hosted by a post-main-sequence star, Biotechnosignature 5 could be a combination of the biosignatures and technosignatures produced by a migrating civilization in the outer regions of the planetary system and the biosignatures of bacteria remaining on a planet experiencing a gradual destruction by the radiation of its host star, with some commonalities shared by the bacterial biosignatures produced on the planets in the outer regions of the planetary system and the bacterial biosignatures produced on the planet experiencing destruction by the host star.

To distinguish the biosignatures of 'natural' bacteria surviving on the abandoned planet from the biosignatures of engineered bacteria, models would have to be created and used to estimate the possible survival time of 'natural' bacteria and the possible strength of their biosignature. Models are already developed and used to estimate the duration of Earth's habitability and, therefore, the life span of Earth's biosphere. According to one model, for example, the end of Earth's biosphere would happen long before the Sun becomes a red giant, as the biosphere would collapse due to high temperatures; however, the end of the biosphere would hardly happen sooner than $1.5 \times 10^9$ years (de Sousa Mello and Friaça, 2020). Another model was used to estimate the temperature evolution of Earth over the next $3 \times 10^9$ years; its results suggested that even after the extinction of complex life forms on Earth orbiting the post-main-sequence Sun, single-celled life forms could persist in high-latitude regions of Earth for up to $2.8 \times 10^9$ years from the present (O'Malley-James *et al*., 2013). Similar models could be used for exoplanets orbiting post-main-sequence stars.

Therefore, Biotechnosignature 5 can be also characterized by the bacterial biosignatures detected on exoplanets gradually destroyed by their post-main-sequence host stars, when such bacterial biosignatures last for greater periods of time than those predicted by the models for 'natural' bacteria, indicating that engineered bacteria could produce the bacterial biosignatures. On its own, this characteristics of the biotechnosignature may not be conclusive enough, as extraterrestrial bacteria might naturally evolve to survive for greater periods of time on the planets affected by their post-main-sequence host stars.

## *Biotechnosignature 6: Biologically inspired technologies*

Hypothetical extraterrestrial civilizations may expand their existing biotechnospheres or create new biotechnospheres by incorporating biologically inspired technologies that operate similarly to life forms. Man-made biologically inspired technologies can offer insights into possible extraterrestrial biologically inspired technologies as follows.

### *Machine learning*

A growing body of research deals with adaptation of behavioural patterns and social phenomena observed in nature towards efficiently solving computational tasks for a number of domains such as energy, climate, health and many others (Del Ser *et al*., 2019). In addition to many machine learning systems relying on the concept of the neural networks of the human brain, some developments in the field of AI involve AI algorithms imitating microbial intelligence; for example, swarm intelligence algorithms, a subset of AI algorithms, are biologically inspired optimization algorithms; one of them is a bacterial foraging optimization (BFO) algorithm mainly simulating the behaviours of *Escherichia coli* searching for nutrients (Tang, *et al*., 2021).

### *Computing technologies*

Neuromorphic computing, a method of computer engineering that models elements of computing systems after systems in the human brain and nervous system, is one of the examples of biologically inspired computing technologies (Furber, 2016; Marković *et al*., 2020). Computing systems were also compared with the processes in single-celled organisms described in terms of computational processes (Bray, 2009). Bray referred to an individual cell as a robot made of biological materials and compared bacterium to a parallel distributed processing (PDP) network; Bray described wetware as the sum of all the information-rich molecular processes inside a living cell, exhibiting resemblance and distinction when compared with the hardware of electronic devices and the software encoding memories and operating instructions (Bray, 2009). Some limitations of Bray's approach were pointed out as follows: (i) there are no wires connecting enzymes in a pathway in the cell, and the cell relies on diffusion and compartmentalization in the form of organelles; (ii) because cellular circuitry is noisy, its outcome can be difficult to predict and (iii) the molecular circuits of a cell are malleable and depend on the environmental conditions (DeMare, 2011).

If some hypothetical advanced extraterrestrials preferred computing systems with a capacity for adaptive change, they could create computing technologies imitating the computing powers of single-celled life forms. Possible commonalities of single-celled life forms in the Galaxy could help human scientists recognize the biologically inspired nature of such computing technologies if they were discovered *in situ*. On the other hand, finding associations of the properties of hypothetical extraterrestrial computing technologies with the 'brain' of advanced extraterrestrial species could be challenging because of all the possible differences between the human brain and the information-processing organs of advanced extraterrestrials.

*Robotics*

Bio-inspired robotics is another broad research area (Peyer *et al.*, 2013; Iida and Ijspeert, 2016; Kim *et al.*, 2018; Wang et al., 2022). Biological inspiration, for example, could be used to design self-replicating probes resembling hypothetical Von Neumann probes (Bracewell, 1960; Freitas and Zachary, 1981; Matloff, 2022). Namely, an efficient approach to the design of Von Neumann probes could be a small payload, which then could build what is required *in situ*; biologically inspired examples for this approach include *Vibrio comma* as the smallest replicator in a general environment and the slightly heavier *E. coli* being a very robust replicator; so a final replicator could have a mass of 30 g, including the AI and the manipulator arms (Armstrong and Sandberg, 2013). To prevent the grey goo problem, a biologically inspired approach based on telomeres can be used to investigate how the number of offspring spawned by self-replicating may be controlled at a genetic level (Ellery, 2022b).

If extraterrestrial microorganisms were discovered beyond Earth in the Solar System, their microbial intelligence and their biological properties could provide novel ideas for creation of biologically inspired technologies and AI algorithms. If microbial intelligence existed in other planetary systems and had commonalities with microbial intelligence existing in the Solar System, then hypothetical advanced civilizations inhabiting the other planetary systems could use similar biological ideas to develop similar biologically inspired robotic systems, computing systems, machine learning and AI algorithms. Therefore, studies of microbial intelligence existing on Earth and beyond Earth (if microorganisms were discovered beyond Earth in the Solar System) could provide insights into hypothetical biologically inspired technologies and AI algorithms that exosolar extraterrestrial civilizations could create.

Accordingly, Biotechnosignature 6 could be in the form of the technosignatures and artefacts of biologically inspired extraterrestrial technologies and machine learning algorithms if such were discovered and their biological inspirations could be recognizable based on what we know about life in the Solar System.

Because almost all robotic systems made by human civilizations are inspired by biological systems (Iida and Ijspeert, 2016; Wang *et al.*, 2022), an assumption can be made that many properties of extraterrestrial robotic systems may be inspired by the properties of life forms known to extraterrestrial civilizations. So that the properties of extraterrestrial robotic systems, if discovered, could provide hints on the properties of extraterrestrial life forms, even if these properties were very different from the properties of life forms on Earth. In this way, the biotechnosignature of the extraterrestrial robotic systems could also be the biosignature of the biological species that created the robotic systems or the biosignature of other life forms known to such a biological species.

### *Biotechnosignature 7: Terraformation of exoplanets and exomoons*

Tools and methods developed to create biotechnospheres that would help to recover Earth's environments and ecosystems could be applied in the design of biotechnospheres that would support the ecosystems of habitats for humans beyond Earth and enable terraformation of other planets (Solé *et al.*, 2018). Terraforming experimentations on Mars, for example, could involve the development of a biotechnosphere comprising synthesized microorganisms monitored by technologies. These terraforming experimentations could suggest Biotechnosignature 7 in the form of observables of terraforming operations in planetary systems. For example, an advanced extraterrestrial civilization performing multiplanetary terraforming in a planetary system hosted by a main-sequence-star, could produce a biotechnosignature in the form of biosignatures, technosignatures and characteristics of the modified environments detected on different planets and (or) moons and, yet, demonstrating some commonalities.

### *Biotechnosignature 8: Extinct biotechnospheres on Earth*

Technosignatures can outlive civilizations that created them (Carrigan, 2012; Davies, 2012; Stevens *et al*., 2016; Balbi and Ćirković, 2021), and so can biotechnosignatures. If an advanced civilization would become extinct or abandon the worlds where it created biotechnospheres, the biotechnospheres could continue to function and produce biotechnosignatures for some time, but their technologies could eventually stop functioning. The discovery of bacteria that were part of such extinct biotechnospheres could blur the line between astrobiology, interstellar archeology and the search for the artefacts of extraterrestrial civilizations. That is, if the technologies of an extinct biotechnosphere on an exoplanet were not detected at interstellar distances or were not discovered *in situ*, then the engineered microbes of the extinct biotechnosphere or their descendants could be mistakenly identified as naturally emerging microbes. However, *in-situ* studies could potentially determine or suggest if some of microbes currently existing on Earth or potentially discovered by future space missions on other planets, moons and asteroids could be descendants of the microbes that were part of ancient biotechnospheres (if such biotechnospheres existed).

The possibility of another civilization creating engineered microbes on Earth was previously discussed in a possible scenario of an alien expedition, probe or colony using biotechnology to modify terrestrial genomes for various practical purposes, and it was proposed that evidence of the modifications could exist in terrestrial genomes to this day, hidden in genetic data (Davies, 2012). According to another scenario, which is a variant of Crick's directed panspermia hypothesis (Crick and Orgel, 1973), extraterrestrials could create an artificial 'shadow biosphere' (i.e. Life 2.0) in the form of microorganisms with biochemistry different from that discovered so far on Earth, and remnants of the shadow biosphere could exist unrecognized on Earth (Davies and Lineweaver, 2005; Davies *et al*., 2009; Davies, 2012). It was proposed to search for such weird microorganisms in the unsampled niches on Earth (Davies, 2012). In addition to the idea of advanced extraterrestrials visiting Earth in the distant past, a few studies posited and discussed the possibility of the rise and existence of technologically advanced civilizations on ancient Earth, Mars or Venus (Wright, 2018; Schmidt and Frank, 2019).

The questions about the possible existence of previous advanced civilizations in the Solar System and the possible existence of an ancient shadow biosphere of Earth comprising engineered microbes (Davies and Lineweaver, 2005; Davies *et al*., 2009; Davies, 2012) is extended here to pose the following question: Is it possible to find evidence of any hypothetical ancient advanced civilization existing on Earth and creating a planetary biotechnosphere or local biotechnospheres on Earth in the distant past?

Potentially, evidence could come from the descendants of the bacteria that were part of a biotechnosphere created by another advanced civilization on Earth in the distant past. The bacteria of the ancient biotechnosphere would not necessarily be strikingly 'weird', as the civilization could modify bacteria already existing on Earth to make them usable as part of the biotechnosphere. The descendants of these bacteria could survive to our times and have commonalities with other bacteria currently populating Earth. At the same time, the modern descendants of the engineered ancient bacteria could preserve in themselves a combination of properties and abilities related to the role that their ancestral bacteria played in the ancient biotechnosphere. Biotechnosignature 8 would exist in the form of a combination of the properties and abilities inherited by some bacteria from their ancestral bacteria that were part of ancient biotechnospheres on Earth, if such biotechnospheres existed. These properties and abilities are described as follows.

#### *Specialization*

Because bacteria would perform specific tasks under specific conditions in the ancient biotechnospheres, they would be engineered to have some sort of specialization such as ecological specialization and (or) metabolic specialization. It could be argued that another civilization could prefer to use bacteria-generalists capable of switching between different tasks and demonstrating broad environmental tolerances. However, hypothetical civilizations could prefer to avoid an extensive use of bacteria-generalists in their biotechnospheres on planets with existing biodiversity because more complex engineered biological entities could be prone to a greater number of malfunctions.

### *Speciation*

Horizontal gene transfer is so pervasive among bacteria that it can reduce genetic isolation between bacterial populations (Caro-Quintero and Konstantinidis, 2012). The designers of the ancient biotechnospheres would want to prevent a transfer of genetic information between 'natural' bacteria of the planet and the engineered bacteria to preserve the properties and composition of the biosphere and the biotechnospheres. Therefore, bacteria would have to be engineered to show phenotypic cohesion and to have bacterial genotype preserved with the help of various mechanisms preventing horizontal gene transfer. Some studies in the field of biotechnology and synthetic biology, for example, already search for different ways of creating barriers that prevent dissemination of genes among bacteria via horizontal gene transfer (e.g. Corvaglia *et al*., 2010).

Additionally, the engineered bacteria would need to have their bacterial genotype preserved in the presence of viruses. The most dominant form of viruses in the virosphere of Earth are bacteriophages, which suggests that the 'natural' bacteriosphere and the virosphere are structurally coupled (Moelling and Broecker, 2019). Comparative genomics recognized that the chromosomes from bacteria and their viruses (bacteriophages) are coevolving, and studies of bacterial pathogens confirmed this process; namely, the majority of bacterial pathogens contain prophages or phage remnants integrated into the bacterial DNA, and many prophages from bacterial pathogens encode virulence factors (Brüssow *et al*., 2004). To avoid this, the bacteria engineered as part of the biotechnospheres would have to be created with the means to decouple and isolate themselves from the influence of the virosphere.

### *Ability to switch between dormant and active states and perform other actions in response to artificially created external stimuli*

Dormancy is used by microorganisms as a bet-hedging strategy. It has important consequences for ecosystem-level processes, and it may help to explain numerous ecological phenomena in microbial systems (Lennon and Jones, 2011). If bacterial communities and bacterial consortia were engineered to function as part of an ancient biotechnosphere, there would be a need to control the rate of their metabolic and other activity so that their collective activity would properly assist with regulating the planetary environment and ecosystems. This could be achieved, for example, by modifying the number of engineered bacteria acting in the biotechnosphere at any given time.

For this purpose, the bacteria could be genetically modified or synthesized to respond to certain external stimuli by shifting from dormant to active state and from active to dormant state. For example, a study of *Bacillus subtilis* spores demonstrated that during dormancy, these spores gradually release their stored electrochemical potential to integrate extracellular information over time, and the decision to exit dormancy can be modulated by genetically and chemically targeting potassium ion flux (Kikuchi *et al*., 2022). Because the objective of these transitions would be to help the biotechnosphere to sustain the planetary environment and ecosystems, the transitions would be initiated by the external stimuli, and they would not necessarily serve the needs of the engineered bacteria. Alternatively, the bacteria could be designed to remain active, but change their metabolic and other activity to some extent in response to external stimuli.

The descendants of such bacteria could also demonstrate similar properties, as in being 'designed' to response to certain external artificial stimuli by readily shifting between their dormant and active states or changing their behaviour in some other ways (e.g. changing their motility).

*Collective properties of bacterial communities and bacterial consortia indicating that they could be engineered to be part of a bigger system*

Microbes have been engineered to address a variety of biotechnological applications, including biosynthesis and bioremediation; a promising direction in these developments involves harnessing the power of designer microbial consortia comprising multiple populations with well-defined interactions, as consortia can complete tasks that are difficult or possibly impossible to complete using monocultures (Tsoi *et al*., 2019). Accordingly, there is a call for researchers to apply microbial ecology to create the environmental biotechnologies to help with advancements in the engineering of microbial communities (Fowler and Curtis, 2023), and there is also an ongoing discussion of how systems biology approaches can be relevant to microbial ecology (Otwell *et al*., 2018). Synthetic biology makes it possible to examine cooperation in microbial systems; manipulate microbial strategies within a population; obtain insights into heterotypic partnerships, including cross-feeding interactions and spatial self-organization; advance towards engineering complex microbial ecosystems for industrial, bioremediation and therapeutic purposes (Rodríguez Amor and Dal Bello, 2019).

These studies could exemplify research in the direction of development of biotechnospheres on Earth. Methods and approaches used in such current studies and future studies in the fields of microbial ecology, systems biology, synthetic biology and biotechnology may also be applied in the future to seek artefacts of ancient biotechnospheres. Specifically, they could be applied to search for evidence of synthetic cooperation in currently existing microbial communities, as well as evidence of responsiveness of such communities to artificial external stimuli and the ability of such communities to produce signals that could be read by technologies. If any microbial communities were discovered to have these unusual properties, they could be studied as potential descendants of the microbial communities engineered in the past for applications in the ancient biotechnospheres.

Currently, the distinction in categorization of microbial communities as 'engineered' or 'natural' becomes blurred in some cases on Earth. For example, wastewater treatment facilities and algal ponds are systems engineered for a specific goal (e.g. water purification, biofuel production or both) but they are subject to environmental variations and influx of invasive species (Song *et al*., 2015). A civilization creating a mature biotechnosphere would want to prevent mixing and clashing of 'engineered' and 'natural' microbial communities. The communities of the descendants of the microbes engineered as part of the ancient biotechnosphere could also demonstrate a trend of avoidance of mixing and clashing with other microbial communities.

*Traces of another civilization's technological intervention with the genetic code of the bacteria*

It was proposed that evidence of the modifications of bacterial genetic code done by another civilization could exist in terrestrial genomes to this day, hidden in genetic data (Davies, 2012). Therefore, it could be anticipated that the genetic code of the bacteria engineered to be part of the biotechnosphere could include traces of another civilization's technological intervention, which could be potentially preserved and identified in the descendants of these bacteria. One of these genetic manipulations could aim to provide the bacteria with advanced mechanisms of self-repair of mutations.

### *Biotechnosignature 9: Biotechnospheres in the Solar System beyond Earth*

Exosolar advanced civilizations could visit the Solar System or send technologies to the Solar System, where the technologies would produce technosignatures or continue exist as extraterrestrial artefacts (Freitas and Valdes, 1985; Arkhipov, 1995, 1998a, 1998b; Haqq-Misra and Kopparapu, 2012; Davies and Wagner, 2013; Benford, 2019, 2021; Romanovskaya, 2022). A few hypotheses posited that technologically advanced civilizations could emerge on ancient Earth, Mars or Venus and leave technosignatures on Earth and elsewhere in the Solar System (Wright, 2018; Schmidt and Frank, 2019).

Whether any hypothetical extrasolar advanced civilizations visited the Solar System in the past or some non-human civilizations emerged in the Solar System long before the rise of the human intelligence, those civilizations could establish biotechnospheres in the Solar System beyond Earth to extract resources from asteroids and other objects of the Solar System and (or) to terraform selected objects of the Solar System (e.g. Venus, Mars or Jupiter's moon Europa).

Cockell discussed previously published data from experiments in three areas of geomicrobiology that could be applied to space settlement: soil formation from extraterrestrial regolith, biological extraction of economically important elements from rocks and biological solidification of ground on other planetary surfaces (Cockell, 2011). Cockell used the data to propose attributes that could be introduced into engineered microbes in these applications, as well as a set of 'core' attributes that could be introduced into any microorganisms used in space geomicrobiology; the set of the 'core' attributes would include: (1) rapid growth rate combined with tolerance of extreme extraterrestrial environments; (2) tolerance of metals found in regolith rocks; (3) ability to fix nitrogen; (4) ability to grow under a wide diversity of chemical and physical conditions; (5) the minimal element needs to allow for growth in nutrient-deprived rock environments and (6) robust resting states (i.e. spores) for long-term storage in planetary stations or when transported between planets (Cockell, 2011).

From the point of view of this study, Cockell investigated how humankind could establish localized biotechnospheres beyond Earth in the Solar System. The properties of the microbes engineered as part of these biotechnospheres would include the set of the 'core' attributes proposed by Cockell (2011) combined with the properties of engineered microbes proposed here as Technosignature 8 (i.e. shifting between a dormant and active state and performing other tasks in response to artificially created external stimuli; potential signs of artificial interference with the genetic code of the bacteria; avoidance of horizontal gene transfer and immunity to the influence of viruses).

Therefore, Biotechnosignature 9 could be in the form of: (i) a set of the properties of microbes engineered to function as part of biotechnospheres beyond Earth, where these properties would include rapid growth rate combined with tolerance of extreme extraterrestrial environments, tolerance of metals and other substances found in regolith rocks, ability to fix nitrogen, ability to grow under a wide diversity of chemical and physical conditions, minimal element needs to allow for growth in nutrient-deprived environments, robust resting states, ability to shift between dormant and active state and perform other tasks in response to artificially created external stimuli, presence of artificial interference with the genetic code, avoidance of bacterial wars, avoidance of horizontal gene transfer and immunity to the influence of viruses and (or) (ii) artefacts of the operation of such microbes.

This biotechnosignature could also exist in the form of (i) bacterial descendants of the bacteria used in ancient biotechnospheres if such bacteria would continue to exist in subsurface microbial ecosystems or subsurface 'deposits' of extraterrestrial bacterial spores on Mars and other objects of the Solar System and (ii) modifications of regolith and rocks on cosmic objects made by the bacteria-descendants and displaying commonalities with applications of biomining.

### *Biotechnosignature 10: Biotechnospheres on free-floating cosmic objects*

Future robotic space missions could search *in situ* for biotechnosignatures of extant or extinct biotechnospheres on free-floating cosmic objects (e.g. free-floating planets, interstellar comets and interstellar asteroids) passing through the Solar System. Biotechnosignature 10 that they could host could be similar to Biotechnosignature 8 and Biotechnosignature 9.

Some free-floating planets may have habitable conditions, host simple life forms and deliver them to planetary systems (Stevenson, 1999; Abbot and Switzer, 2011; Badescu, 2011; Schulze-Makuch and Fairén, 2021); some free-floating planets can be potentially habitable Earth-sized free-floating planets with subsurface oceans (Abbot and Switzer, 2011). The existence of exomoons orbiting free-floating planets was theoretically predicted and a study proposed that, under certain conditions and assuming stable orbital parameters, liquid water could exist on the surface of such exomoons, and the amount of water could be sufficient to provide habitable conditions and host primordial life (Ávila *et al.*, 2021). Lingam and Loeb briefly discussed some of these studies of the potential habitability of free-floating planets and concluded that it follows from these studies that 'it is evident that life in the Universe has a vast range of niches that it could occupy, and worlds with subsurface oceans under ice envelopes constitute an important category'; Lingam and Loeb also discussed how the primordial Earth might have retained a global subsurface ocean if Earth had become a free-floating planet (Lingam and Loeb, 2019).

Some extraterrestrial civilizations, if they exist, may use free-floating planets as a means of interstellar transportations or they may send machines to ride free-floating planets (Romanovskaya, 2022). These civilizations could create subterranean biotechnospheres on free-floating planets and biotechnospheres in the oceans of free-floating planets and their exomoons (e.g. to extract resources). If these civilizations become extinct or abandon their free-floating planet, the free-floating planets (and, potentially, their exomoons) could continue to carry the biotechnospheres or the fragments and artefacts of the biotechnospheres (and, hypothetically, become involved in lithopanspermia).

Interstellar asteroids could also harbour technosignatures and biotechnosignatures of mining and biomining that could be done by advanced civilizations or intelligent machines (i.e. machines similar to Von Neumann probes) when the asteroids were gravitationally bound to planetary systems or after they became free-floating asteroids. Namely, some asteroids originate in the inner regions of the planetary systems hosted by main-sequence stars and become ejected from the inner regions to the Oort Clouds of their planetary systems (Shannon *et al*., 2015). Hypothetically, mining and biomining operations could take place on these Oort-cloud asteroids in their Oort Clouds. The Oort-cloud asteroids could later become ejected from their planetary systems by their post-main-sequence stars (Veras *et al*., 2011; Veras and Wyatt, 2012).

The following scenarios describe this possibility: (1) to survive the post-main-sequence evolution of their host stars, advanced civilizations would migrate from their planetary system's inner regions to their Oort Clouds (Romanovskaya, 2022) and use mining and biomining to extract resources from asteroids in the Oort Cloud; (2) to escape existential threats, advanced civilizations could migrate from their home planetary system to the Oort Clouds of other planetary systems (Romanovskaya, 2022), where the civilizations could use mining and biomining to extract resources from asteroids in the Oort Cloud; (3) interstellar travellers (e.g. Von Neumann probes) could visit a young planetary system and perform mining and biomining operation on its asteroids and (4) post-biological entities (e.g. intelligent machines) could migrate to the Oort Cloud of their home planetary system and use mining or biomining to extract resources from Oort-cloud asteroids (this hypothetical scenario is based Ćirković and Bradbury's migration hypothesis: Ćirković and Bradbury, 2006).

## The Cosmic Descendants hypothesis

The migration hypothesis proposed by Ćirković and Bradbury posits that post-biological entities (e.g. non-biological computing entities) would migrate outward from their original location in the Galaxy towards the outer regions of the Galaxy where temperature is low enough to increase their computing efficiency, as computation becomes more efficient when the temperature of the heat reservoir in contact with the computing technologies is lower, and the most efficient heat reservoirs are the regions of the Universe located far from energy sources (e.g. stars) (Ćirković and Bradbury, 2006). In the migration hypothesis, Ćirković and Bradbury generalized the idea of another type of migration of post-biological entities within the Solar System; namely, Ćirković and Bradbury discussed how post-biological descendants of humankind could prefer low-temperature and volatile-rich outer regions of the Solar System, thus creating 'circumstellar technological zone' that would be different and complementary to the circumstellar habitable zone (Ćirković and Bradbury, 2006).

On the other hand, the Silurian hypothesis involves an examination of the possibility of detecting evidence of a prior industrial civilization in Earth's geologic record, with an assumption that such a civilization could exist on Earth millions of years before humans (Schmidt and Frank, 2019). This possibility is further discussed in this article (i.e. Biotechnosignature 8).

While considering the migration hypothesis (Ćirković and Bradbury, 2006) and the Silurian hypothesis (Schmidt and Frank, 2019) alongside, I propose the Cosmic Descendants hypothesis that posits that if an industrial civilization of non-human biological species emerged on Earth, Mars or Venus in the distant past, as other studies suggested (Wright, 2018; Schmidt and Frank, 2019), the civilization could be survived by the bacteria, which the civilization engineered for its biotechnospheres, and the civilization could be survived by intelligent machines it created; namely, the ancient industrial civilization could build biotechnospheres on Earth, Mars or Venus, the biotechnospheres could incorporate engineered bacteria and the descendants of that engineered bacteria could exist on Earth or Mars (and, highly hypothetically, in the atmosphere of Venus, if the artefacts of the biotechnospheres were airborne) till this day, bearing some properties of their engineered bacterial ancestors; if that civilization was survived by the societies of its intelligent machines, the intelligent machines could migrate to the outer regions of the Solar System and survive there until present times. The societies of the intelligent machines (or even their own civilization, if they would organize as a civilization) could use biotechnospheres, for example, for biomining operations to extract resources from comets, asteroids and other objects in the main asteroid belt and in the outer regions of the Solar System.

The location and properties of the biotechnospheres built by the civilization of intelligent machines would depend on its location relative to the heliosphere (i.e. inside the heliosphere or outside of the heliosphere) because the heliosphere to a significant degree shields the region of the Solar System, which it encompasses, from Galactic Cosmic Rays (GCRs) that are harmful to life and electronics (Zeitlin *et al.*, 2016). The solar wind affects the heliosphere's size and shape, and so does the Sun's motion through the local interstellar medium because it compresses the heliosphere at the front and drags it out into a tail at the back. As a result, the distance of the heliosphere's leading edge from the Sun is far less than that to the end of the heliotail; NASA's Voyager 1 crossed the heliopause in the direction of its leading edge in mid-2012 at a distance of about 122 AU from the Sun (Stone *et al.*, 2013), while Voyager 2 encountered the heliopause in late 2018 at a distance of 119 AU (Stone *et al.*, 2019).

The inner edge of the Oort Cloud is estimated to be at $\sim 2 \times 10^3$ AU (Fouchard *et al.*, 2017), which means that the heliosphere does not protect the Oort Cloud from GCRs and the Oort-cloud comets and asteroids are affected by GCRs. GCRs penetration depths into solid matter depend on their energies, $E$, are as follows: for $E \lesssim 0.1$ GeV, the penetration depths are <0.1 m (Cooper *et al.*, 2004; Gronoff *et al.*, 2020a) and for $E \gtrsim 1$ GeV, protons and alpha particles have penetration depths in ice ~1 to 10 m (Jewitt and Seligman, 2022). If the civilization of intelligent machines resided in the Oort Cloud, its biotechnospheres and biomining operations could be limited to the interiors of the Oort-cloud asteroids, probably several metres below the asteroids' surface.

The civilization of intelligent machines could include non-biological technologies only. Alternatively, the civilization of intelligent machines could use both biological and non-biological components in the design of its machines and other technologies. This possibility should be acknowledged because scientists on Earth are already working on developing such technologies. Examples include memristive connections linking silicon neurons and brain neurons of the rat hippocampus (Serb *et al.*, 2020); *in vitro* neural networks from human or rodent origins integrated with *in silico* computing (Kagan *et al.*, 2022); synthetic gene networks constructed to emulate digital circuits and devices (Friedland *et al.*, 2009) and synthetic gene networks for AI (Nesbeth *et al.*, 2016). Because of the biological components of the machines, the civilization of intelligent machines could prefer to be in the Kuiper Belt and in the inner region of the scattered disk so that the heliosphere would protect them from GCRs. The civilization of intelligent machines could also inhabit or, at least, explore the oceans of Ceres and the moons of Jovian planets for resources. It would likely reside outside the inner Solar System to distance itself from extreme solar events (e.g. coronal mass ejections, solar flares and superflares).

The design of the intelligent machines would also affect their migration patterns in the Galaxy. The civilization of intelligent machines built with non-biological components could follow the migration scenario proposed in the migration hypothesis (Ćirković and Bradbury, 2006) and migrate towards the outer regions of the Galaxy where temperature is low enough to improve their computing efficiency and where they could distance themselves from destructive cosmic phenomena (e.g. supernovae) (Ćirković and Bradbury, 2006). The civilization of intelligent machines incorporating biological and non-biological would have to account for the needs of the machines' biological components, and so it could remain in the Solar System for greater periods of time. Highly hypothetically, such a hypothetical civilization could be the first and the only line of defense of the Solar System from hostile exosolar visitors that are so frequently pictured in many works of science fiction.

Possible technosignatures and biotechnosignatures of such a civilization of intelligent machines are proposed as follows.

### *Technological artefacts and technological activities of intelligent machines in the outer Solar System*

When discussing the search for artefacts of hypothetical extinct prior civilizations of the Solar System, Wright proposed that photometry and spectra of asteroids, comets, and Kuiper Belt Objects could reveal albedo, shape, rotational, compositional or other anomalies because such objects could host artefacts, or they could be artefacts (Wright, 2018). Technosignatures representing these anomalies could also be produced by inactive abandoned technologies or currently active technologies of the civilization of intelligent machines, which survived the prior civilization of the Solar System that created the machines. For example, the intelligent machines' technosignature could be in the form of infrared and other electromagnetic radiation produced by mining operations on asteroids. Forgan and Elvis suggested that the mining operations of advanced civilizations could produce technosignatures in debris discs in the form of electromagnetic radiation detectable by astronomical tools (Forgan and Elvis, 2011), the same consideration could be applied to the technosignature of machines mining asteroids in the Solar System.

### *Lurkers*

Benford discussed how hypothetical exosolar advanced civilizations could send robotic surveillance probes (Lurkers) to the Solar System, place them on nearby co-orbital objects to observe Earth and have the Lurkers sending surveillance data back to their origin; one of their technosignatures could be the probes themselves, if they were discovered *in situ* (Benford, 2019). Technosignatures in the form of Lurkers could also be produced by the intelligent machines that survived the prior civilization of the Solar System if they would place the Lurkers to survey Earth and humankind.

### *Surveillance probes on the Moon*

Surveillance technology of hypothetical exosolar advanced civilizations could be located on the Moon, and potential existence of extraterrestrial artefacts on the Moon was discussed in other studies (e.g. Arkhipov, 1995, 1998a, 1998b; Davies and Wagner, 2013). The intelligent machines could also place their surveillance probes on the Moon.

### *Bacterial descendants of ancient bacteria that were part of biotechnospheres designed to terraform or sustain planetary environments (Biotechnosignature 8)*

If a hypothetical prior non-human civilization could be advanced enough to create biotechnospheres on Earth, Mars or Venus in the distant past, the civilization could also be advanced enough to build spacefaring intelligent machines that could later migrate to the outer Solar System. The intelligent machines could establish biotechnospheres elsewhere in the Solar System and they could use the same engineered bacteria in their biotechnospheres that the ancient civilization used in the biotechnospheres established on Earth, Venus or Mars. Although the engineered bacteria would be modified to a certain extent for the new environments (e.g. when used to terraform the subsurface ocean of Ceres, the subsurface oceans of the moons of Jovian planets), they would still share commonalities with the bacteria of the ancient biotechnospheres established by the prior civilization on Earth, Venus or Mars.

Consequently, if bacterial descendants of the bacteria that were part of the ancient biotechnospheres built on Earth, Venus or Mars were discovered, their properties could be similar to the properties of the engineered bacteria used by the intelligent machines in their biotechnospheres on other objects of the Solar System (e.g. in the subsurface ocean of Ceres, the subsurface oceans of the moons of Jovian planets). The existence of such bacteria and their commonalities would serve as a biotechnosignature. Considerations of the search for bacteria on Venus mentioned here are pertaining to the airborne artefacts of biotechnospheres if such artefacts could survive in the atmosphere of Venus.

The possibility of this biotechnosignature would suggest searching for potentially engineered bacteria on Earth and Mars, in the Venusian atmosphere, in the ocean of Ceres, the oceans of the moons of Jovian planets and in the water plumes of Saturn's moon Enceladus.

### *Solar probes as part of biotechnospheres (as discussed for Biotechnosignature 3)*

If the civilization of intelligent technologies would reside inside the heliosphere (e.g. in the Kuiper Belt), then its technologies and biomining operations would be protected from GCRs by the heliosphere, but they could be impacted by extreme solar activity (e.g. solar flares and solar superflares). Even though the impact of the extreme solar activity would be less significant in the Kuiper Belt than that experienced at 1 AU from the Sun, the civilization of intelligent technologies could still experience the impact and it could have to place space probes in the inner Solar System that would survey the Sun and provide warnings about the magnetic activity of the Sun, the dynamics of the solar wind and upcoming extreme solar events. The civilization could place the probes on Mercury, stable Venus co-orbital asteroids, Aten asteroids or Apollo asteroids, so that they would monitor the Sun in close proximity. It could also place solar probes on the Moon, Mars and in the main asteroid belt. This is because in addition to the studies of the Sun producing solar wind, flares, superflares and coronal mass ejections, the civilization would also need to know how the solar particles, radiation and magnetic fields would propagate through the Solar System.

Because the probes would gather data about the Sun necessary for the protection of the civilization of intelligent machines and its biotechnospheres, the probes could be considered as technologies belonging to the biotechnospheres and, therefore, representing a biotechnosignature of that civilization (this is also discussed as part of Biotechnosignature 3).

### *Biologically inspired technologies and machines composed of biological and non-biological components (Biotechnosignature 6)*

*In-situ* studies of asteroids, Ceres, moons of Jovian planets and other objects of the Solar System could include the search for artefacts of technologies, including biologically inspired technologies that could be used by the intelligent machines. Societies of intelligent machines composed of biological and non-biological components could be considered themselves as biotechnospheres, as they would combine biology and technology working together towards common goals. The observables of these machines functioning could be described as biotechnosignatures.

### *Artefacts of geomicrobiology applied to space settlements and exploration sites (Biotechnosignature 9)*

Artefacts of geomicrobiology applied by the intelligent machines to their space settlements and exploration sites could serve as their biotechnosignature, this type of biotechnosignature is discussed as Biotechnosignature 9. The possibility of this biotechnosignature could suggest searching for potentially engineered bacteria on Earth and Mars, on Ceres, in the oceans of the moons of Jovian planets and in the water plumes of Saturn's moon Enceladus.

## On the possible transient presence of artefacts of extraterrestrial technologies and biotechnospheres in the Solar System

### *Interstellar asteroids*

Hypothetical advanced extraterrestrial civilizations may place technologies and build subsurface biotechnospheres on Oort-cloud asteroids of their planetary systems (e.g. to perform biomining on the asteroids). The Oort-cloud asteroids can become ejected into interstellar space by their host stars when the stars undergo post-main-sequence evolution (Veras *et al*., 2011; Veras and Wyatt, 2012). If any of these asteroids would travel through the Solar System, then the technosignatures, biotechnospheres or their artefacts existing on the surface and in the interior of the asteroids would have transient presence in the Solar System for the duration of the passage.

The search for technosignatures and biotechnosignatures on the passing interstellar asteroids could be included in the scientific studies of such asteroids with an understanding that the probability of discovering technosignatures and biotechnosignatures on interstellar interlopers (i.e., interstellar comets and interstellar asteroids are described as interstellar interlopers when they are observed passing through the Solar System) as well as the probability of discovering any interstellar asteroid to be an extraterrestrial interstellar spacecraft is extremely low when different possible sources of interstellar comets and interstellar asteroids considered. For example, minor bodies and planetesimals can become interstellar asteroids after they are ejected into interstellar space by the processes of planetary formation and orbital migration of giant planets (Duncan *et al*., 1987; Charnoz and Morbidelli, 2003; Cook *et al*., 2016), and by interactions of young stars in open cluster (Hands *et al*., 2019). It was estimated that $\sim 10^4$ interstellar objects can be located closer to the Sun than Neptune (i.e. distance $\lesssim$30 AU) at any moment of time; with a Solar System crossing time $\sim$10 years, the flux of interstellar interlopers into the planetary region of the Solar System would be $\sim 10^3$ year$^{-1}$ (Jewitt and Seligman, 2022); about 50 interstellar objects >50 m in size could be present in the Solar System in a sphere with a radius of 50 AU at any given time (Borisov and Shustov, 2021).

Whereas young and 'still-under-construction' planetary systems and young open stellar clusters can produce a great number of interstellar comets and asteroids, this discussion of interstellar asteroids travelling through the Solar System is first of all concerned with the interstellar asteroids that originally existed in 'mature' planetary systems, where life would have time to emerge and evolve, and industrial civilizations could have time to become spacefaring civilizations and produce technosignatures and biotechnosignatures on the asteroids. Simulations of the formation of the Oort Cloud of the Solar System demonstrated that $\sim 8 \times 10^9$ of the small bodies in the Oort Cloud are ice-free rock-iron asteroids that formed within 2.5 AU of the Sun (Shannon *et al*., 2015). During its post-main-sequence phase, the Sun may dynamically eject Oort-cloud objects, including these Oort-cloud asteroids, from the Solar System (Veras *et al*., 2011). Other stars of a mass of 1–7 times solar mass could do the same (Veras *et al*., 2011), so that few extrasolar Oort Clouds could survive post-main-sequence evolution intact (Veras and Wyatt, 2012). Therefore, if extraterrestrial civilizations or intelligent machines performed mining or biomining on some asteroids in their Oort clouds and the asteroids were later ejected into interstellar space, the asteroids would carry technosignatures, biotechnosignatures or extraterrestrial artefacts (e.g. the remnants of their biotechnospheres).

A 'low-ball' estimate of the total number of interstellar asteroids that currently remain free-floating in the Galactic disk after they were ejected from the Oort Clouds of planetary systems by their host post-main-sequence stars, which later became white dwarfs in the Galactic disk, can be inferred as $N_{ia} = N_{wd} \times N_a$. Here, $N_{wd}$ is the number of white dwarfs belonging to the Galactic disk and $N_a$ is the estimated number of Oort-cloud asteroids ejected from each planetary system in which the host star underwent its post-main-sequence evolution and now exists as a white dwarf. The Galaxy hosts approximately $10^{10}$ white dwarfs, and the estimated fraction of white dwarfs in the halo is ≈50% of the Galactic white dwarfs (Napiwotzki, 2009). Two assumptions are further made: (i) the average number of Oort-cloud asteroids in each planetary system with its host star destined to become a white dwarf is approximately the same as the number of Oort-cloud asteroids in the Solar System ($\approx 8 \times 10^9$); (ii) for a lower estimate, about 10% of these Oort-cloud asteroids become interstellar asteroids and remain free-floating in the Galactic disk (i.e. not all asteroids may become ejected from the Oort clouds, depending on the size and eccentricity of their orbits, and some asteroids can become ejected from an Oort cloud and yet, experience destruction, become re-captured by other planetary systems or leave the Galactic disk).

With these assumptions, $N_{\mathrm{ia}} \approx 4 \times 10^{18}$, it corresponds to the number density of interstellar asteroids that used to be Oort-cloud asteroids in the Galactic disk and remain free-floating in the Galactic disk, $n \sim 4 \times 10^5$ ly$^{-3}$.

If, hypothetically, $10^8$ advanced civilizations ever existed in the Galactic disk and each civilization would send $10^2$ interstellar spacecraft that could be mistakenly perceived at a distance as interstellar asteroids, then for each such an extraterrestrial interstellar spacecraft in the Galactic disk, there would be more than $4 \times 10^8$ interstellar asteroids that used to be Oort-cloud asteroids and had no traces of extraterrestrial technologies. A similar estimate would exist for interstellar asteroids carrying extraterrestrial technosignatures or biotechnosignatures in the Galactic disk.

### *Free-floating planets*

With an estimate of the number of free-floating planets (with $R \gtrsim 0.3R_{\mathrm{Earth}}$) to be about 30 times the total number of stars, the nearest free-floating planet might be located at a distance that corresponds to the inner Oort cloud of the Solar System (i.e. $\sim 2 \times 10^3 - 2 \times 10^4$ AU) (Lingam and Loeb, 2019). If a free-floating planet hosting a biotechnosphere ever travelled through the Solar System, this would be described as a transient presence of extraterrestrial biotechnosphere in the Solar System. The probability of this event would depend on the probability of the existence of spacefaring extraterrestrial civilizations riding free-floating planets or advanced civilizations sending machines to ride free-floating planets. Hypothetically, lithopanspermia could make the free-floating planet share some of its biotechnosphere with the Solar System.

### *Close stellar flyby*

The closest known flyby of a star to the Solar System was that of the W0720 system, which passed through the Oort cloud of the Solar System approximately $7 \times 10^4$ years ago (Mamajek *et al*., 2015). Another close encounter can be that with Gl 710, with a 95% probability of coming closer than 17 000 AU to the Sun (i.e. passing through the Oort Cloud of the Solar System); the flyby is estimated to occur approximately $1.36 \times 10^6$ years in the future (Bailer-Jones *et al*., 2018). If any object of the planetary system of the close flyby star would carry an extraterrestrial biotechnosphere through the Solar System during the flyby, it would signify a transient presence of their localized biotechnospheres in the Solar System. The probability of this event would depend on the probability of the existence of advanced extraterrestrial civilizations and the probability of such civilizations creating biotechnospheres in the planetary systems of K- and M-type stars, which are the most common main-sequence stars in the Galaxy and the most common main-sequence stars involved in close stellar flybys.

### *Extraterrestrial interstellar spacecraft and probes*

If any extraterrestrial spacecraft or space probes would travel through the Solar System and they would carry biotechnospheres designed to support ecosystems supporting life support, resource utilization inside spacecraft and other purposes, then for the duration of their presence in the Solar System, their localized biotechnospheres would have a transient presence in the Solar System. The probability of this event would depend on the probability of the existence of spacefaring extraterrestrial civilizations using biotechnospheres.

## Conclusions

The concept of planetary intelligence as collective intelligence was applied to investigate the possible evolution of biotechnospheres emerging on the intersection of civilizations' technospheres with planetary biospheres. In mature planetary biotechnospheres, collective intelligence could comprise the intelligence of technologies, the intelligence of life forms, including engineered life forms and, potentially, synthetic intelligence, with all these types of intelligence acting in concert to monitor and preserve planetary biospheres and their biodiversity; to steady planetary environments and to restore them after extinction events; to support space missions and terraformation of cosmic objects; to assist with medical processes, industrial processes, mining, agricultural and food production processes. Biotechnospheres used in space exploration and colonization would become part of cosmic ecosystems and would be likely affected by interactions within the cosmic ecosystems.

Hypothetical advanced civilizations could produce biotechnosignatures (i.e. observables and artefacts of biotechnospheres) in the Solar System, other planetary systems and on interstellar asteroids and free-floating planets. The biotechnosignatures may be in the form of the steadiness of planetary conditions, in some cases accompanied by the long-term steadiness of the biosignatures of complex life forms; unanticipated dynamics of the environments of exoplanets and, in some cases, the unusual dynamics of the biosignatures produced on exoplanets after stellar activity events and events that may cause mass extinctions on exoplanets; unusual persistence of planetary environments and biosignatures on planets strongly affected by their post-main-sequence stars accompanied by the observables of extraterrestrial intelligence migrating to the outer regions of planetary systems; technosignatures of biologically inspired extraterrestrial technologies and machine learning algorithms; certain observables of terraforming operations on exoplanets; the properties inherited by some bacteria from their ancestral bacteria that were part of hypothetical ancient biotechnosphere on Earth or other cosmic objects of the Solar System; artefacts of biomining operations and terraforming operations performed by non-human intelligence in the Solar System, on interstellar asteroids and on free-floating planets. Therefore, some biotechnosignatures could signify the effect of planetary biotechnospheres on the planetary evolution.

Biotechnosignatures could be also in the form of intelligent machines built by hypothetical prior civilizations of non-human species of the Solar System. Such intelligent machines could migrate to the outskirts of the Solar System and beyond, and their design could affect their migration patterns in the Solar System and in the Galaxy. Societies of intelligent machines built with non-biological components could migrate, as proposed in the migration hypothesis (Ćirković and Bradbury, 2006), to the outer regions of the Galaxy. Societies of intelligent machines integrating biological (e.g. *in vitro* neural networks from biological origins, synthetic gene networks, etc.) and non-biological components could remain within the heliosphere for the duration of the main-sequence evolution of the Sun. These machines and their artefacts could reside on the cosmic objects of the Kuiper Belt and the inner regions of the scattered disk, in the subsurface oceans of the moons of Jovian planets and in the subsurface ocean of Ceres.

Transient presence of extraterrestrial technologies and biotechnospheres could occur in the Solar System if interstellar asteroids, free-floating planets or interstellar spacecraft carrying technologies or biotechnospheres would travel through the Solar System or if some object of the planetary system of a close flyby star would carry an extraterrestrial biotechnosphere through the Solar System during the flyby.

A 'low-ball' estimate of the total number of interstellar asteroids that currently remain free-floating in the Galactic disk after they were ejected from the Oort Clouds of planetary systems by their host post-main-sequence stars, which later became white dwarfs in the Galactic disk, is $N_{ia} \approx 4 \times 10^{18}$. Considering this large number, the probability to detect interstellar asteroids carrying biotechnospheres or interstellar spacecraft appearing as interstellar asteroids is very low.

**Financial support.** No funding has been provided for this research.

**Acknowledgements.** I thank the anonymous reviewers for their careful reading of my manuscript.

**Competing interests.** The author reports no conflict of interest.

https://doi.org/10.1017/S1473550423000204 Published online by Cambridge University Press